\documentclass[aps,prb,reprint,amsmath,amssymb,unsortedaddress]{revtex4-2}

\usepackage{graphicx}
\usepackage{dcolumn}
\usepackage{bm}
\usepackage{color}
\definecolor{darkgreen}{RGB}{0,180,0}

\begin{document}

\title{Femtosecond pump-probe absorption edge spectroscopy of cubic GaN}

\affiliation{Institut f\"ur Physik, Otto-von-Guericke-Universit\"at Magdeburg, Universit\"atsplatz 2, 39106 Magdeburg, Germany}
\affiliation{ELI Beamlines/Fyzik´aln´i ´ustav AV ˇCR, v.v.i., Za Radnic´i 835, 25241 Doln´i Bˇrežany, Czech Republic}
\affiliation{Department of Physics, University of Paderborn, Warburger Stra{\ss}e 100, 33098 Paderborn, Germany}

\author{Elias Baron}
\email{elias.baron@ovgu.de}
\author{R\"udiger Goldhahn}
\affiliation{Institut f\"ur Physik, Otto-von-Guericke-Universit\"at Magdeburg, Universit\"atsplatz 2, 39106 Magdeburg, Germany}
\author{Shirly Espinoza}
\author{Martin Zahradn\'{i}k}
\author{Mateusz Rebarz}
\author{Jakob Andreasson}
\affiliation{ELI Beamlines/Fyzik\'{a}ln\'{i} \'{u}stav AV \v{C}R, v.v.i., Za Radnic\'{i} 835, 25241 Doln\'{i} B\v{r}e\v{z}any, Czech Republic}
\author{Michael Deppe}
\author{Donat J. As}
\affiliation{Department of Physics, University of Paderborn, Warburger Stra{\ss}e 100, 33098 Paderborn, Germany}
\author{Martin Feneberg}
\affiliation{Institut f\"ur Physik, Otto-von-Guericke-Universit\"at Magdeburg, Universit\"atsplatz 2, 39106 Magdeburg, Germany}
\date{\today}

\begin{abstract}
Time-dependent femtosecond pump-probe spectroscopic ellipsometry studies on zincblende gallium-nitride (zb-GaN) are performed and analyzed between 2.9-3.7eV. An ultra-fast change of the absorption onset (3.23eV for zb-GaN) is observed by investigating the imaginary part of the dielectric function. The 266nm (4.66eV) pump pulses induce a large free-carrier concentration up to $4\times 10^{20}$cm$^{-3}$, influencing the transition energy between conduction and valence bands due to many-body effects, like band filling and band gap renormalization, up to $\approx$500meV. Additionally, the absorption of the pump-beam creates a free-carrier profile within the 605nm zb-GaN layer. This leads to varying optical properties from sample surface to substrate, which are taken into account by grading analysis for an accurate description of the experimental data. A temporal resolution of 100fs allows in-depth investigations of occurring ultra-fast relaxation and recombination processes. We provide a quantitative description of the free-carrier concentration and absorption onset at the sample surface as a function of relaxation, recombination, and diffusion yielding a characteristic relaxation time of 0.19ps and a recombination time of 26.1ps.
\end{abstract}

\maketitle
\section{Introduction}
\label{sec_intro}

Modern technologies require the application and understanding of faster and faster electronics and optics. From high-speed optical switching \cite{PRL_117_2016,AOM_5_2017} and fast transparent electronics \cite{MT_7_2004,AM_22_2010} to ultra-fast lasers \cite{LSA_4_2015} and computing \cite{NP_12_2016,OPN_27_2016}, the fundamental understanding of ultra-fast phenomena are crucial for the development and deployment. Selection and incorporation of novel materials are no exception. Furthermore, measurement techniques that are able to accurately investigate those effects are necessary. In recent years, time-resolved pump-probe spectroscopic ellipsometry (trSE) proved to be an excellent contestant for this with many possibilities \cite{APL_115_2019,NJP_22_2020,RSI_92_2021,PRR_3_2021}. Spectroscopic ellipsometry is a predestined method for obtaining optical and material properties of semiconductors as well as many other materials, due to its high accuracy and sensitivity. Using the pump-probe approach, systems under strong non-equilibrium conditions can be investigated. In this case, a free-carrier excitation up to $4\times 10^{20}$cm$^{-3}$ are possible, while time-resolutions in the femtosecond regime grant an in-depth analysis of the involved processes. \\
The cubic zincblende phase of group III-nitrides receives increasing interest due to its electronic and optical properties. Especially, zincblende gallium nitride (zb-GaN), and its alloys like InGaN, are promising candidates for closing the so-called green-gap \cite{JDT_3_2007,MST_33_2017,JAP_130_2021}, for high-speed devices \cite{DRM_6_1997} and for qubit applications \cite{JPDAP_54_2021}. Furthermore, the absence of spontaneous and piezoelectric polarization \cite{APL_90_2007,AEM_2_2016} due to the cubic crystal phase are believed to be advantageous over those of the wurtzite phase for certain applications \cite{APL_83_2003,PSSC_9_2012,APL_103_2013,PRB_90_2014_235312}. Although the zincblende phase is metastable, major improvements regarding control and quality of zb-GaN have been reported recently \cite{MST_33_2017,ACSP_5_2018,JAP_124_2018}. In addition, the band structure of zb-GaN is simpler compared to the wurtzite GaN due to higher symmetry \cite{APE_11_2018} and offers a direct band gap of 3.23eV at the $\Gamma$-point of the Brillouin zone (BZ) with no additional local conduction band minima in the vicinity \cite{PRB_77_2008,PRB_84_2011,PRB_85_2012,JAP_71_1992}. This ensures electron-hole pair generation by pump-beam excitation only near the $\Gamma$-point. Furthermore, electron scattering to other conduction band minima within the BZ \cite{APL_115_2019} should be negligible. Other band structure related parameters, like effective masses, have also been studied before \cite{PRB_77_2008,PRB_84_2011,PRM_3_2019}. This should enable zb-GaN to be an promising candidate for understanding the fundamental processes that happen on the femto- and picosecond timescale. \\
The influence of many-body effects like band-filling and band gap renormalization on the absorption onset in semiconductors is widely known \cite{PRM_3_2019,JAP_92_2002,JAP_86_1999,PRB_90_2014_075203,PRB_24_1981}. Understanding these effects is essential for technical applications \cite{PRB_53_1996,PSSB_255_2018}. For highly $n$-type doped materials the Fermi-energy is pushed high into the conduction band. On the other hand, in the experiment discussed here, we create high free electron-hole pair concentrations which should be more comparable to a co-doping situation instead of only $n$-type doping. Therefore, the effect of both high electron and high hole concentration should be considered for an accurate description. Fortunately, the renormalization effect induced by free-hole concentrations seems to be substantially weaker compared to the free-electron contributions \cite{JAP_92_2002}. \\
We report our results from optical investigations on highly excited zb-GaN films in the femto- and picosecond time regime, including the ultra-fast change of the absorption onset visible in the imaginary part of the dielectric function (DF). Furthermore, the pump-beam excitation leads to a free-carrier profile within the sample, which drastically changes the DF at different sample depths. The integration of this profile into the analysis by grading optical properties allows for determining time-dependent transition energies. Considering relaxation, recombination, and diffusion, a model to quantitatively describe the surface free-carrier concentration is derived.


\section{Experimental setup}
\label{sec_experimental}

\begin{figure}[htb]
\includegraphics[angle=0, trim = 0mm 0mm 0mm 0mm, width=0.7\columnwidth]{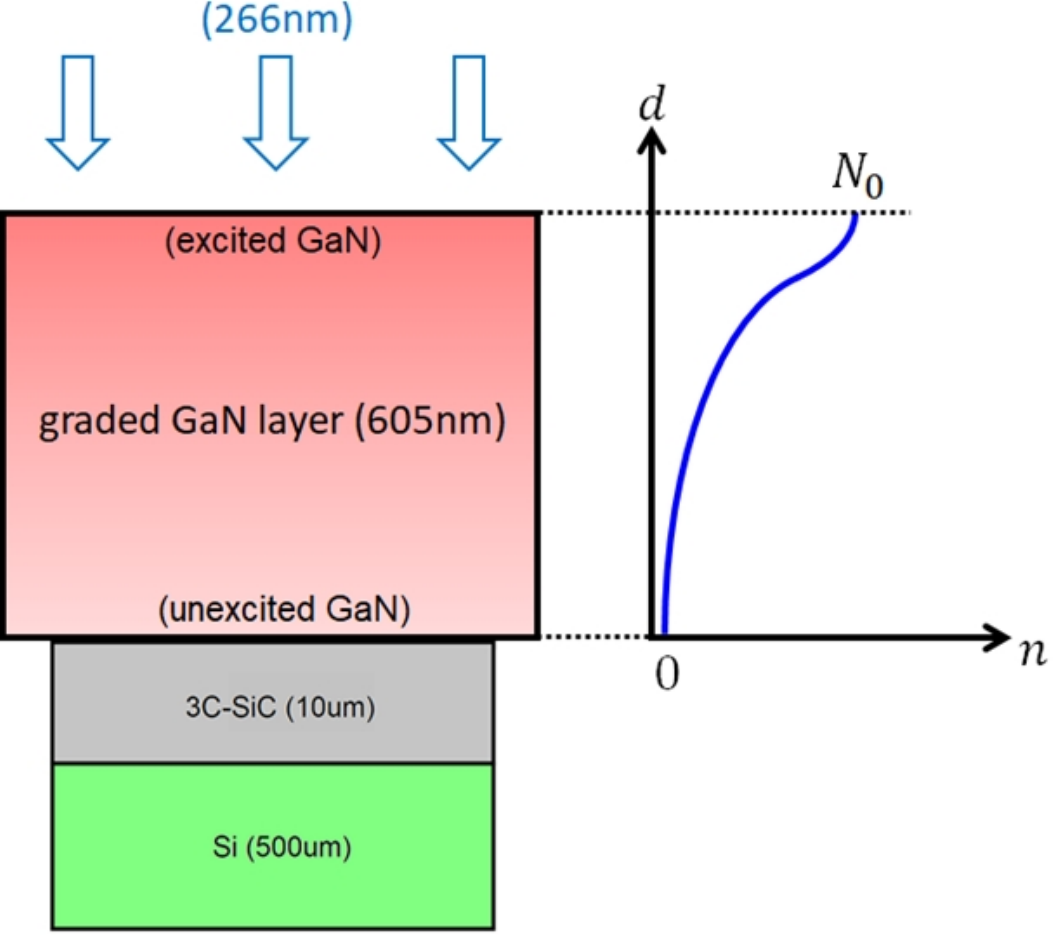}
\caption{Sample structure and sketch of the free-carrier distribution after pump-beam excitation between the excited GaN at the surface and the unexcited GaN at the bottom, considering diffusion and recombination.}
\label{fig_sample}
\end{figure}

In this study, we investigate a 605nm thick zb-GaN thin film grown by plasma-assisted molecular beam epitaxy (MBE) on a 3C-SiC/Si (001) substrate. A schematic sample structure is displayed in Fig. \ref{fig_sample}. The sample was studied by means of steady-state spectroscopic ellipsometry, photoluminescence, Raman spectroscopy, and Hall-measurements earlier. The results of these experiments can be found elsewhere \cite{PRM_3_2019,JAP_125_2019,PSSB_257_2020}.

\begin{figure}[htb]
\includegraphics[angle=0, trim = 0mm 0mm 0mm 0mm, width=1.0\columnwidth]{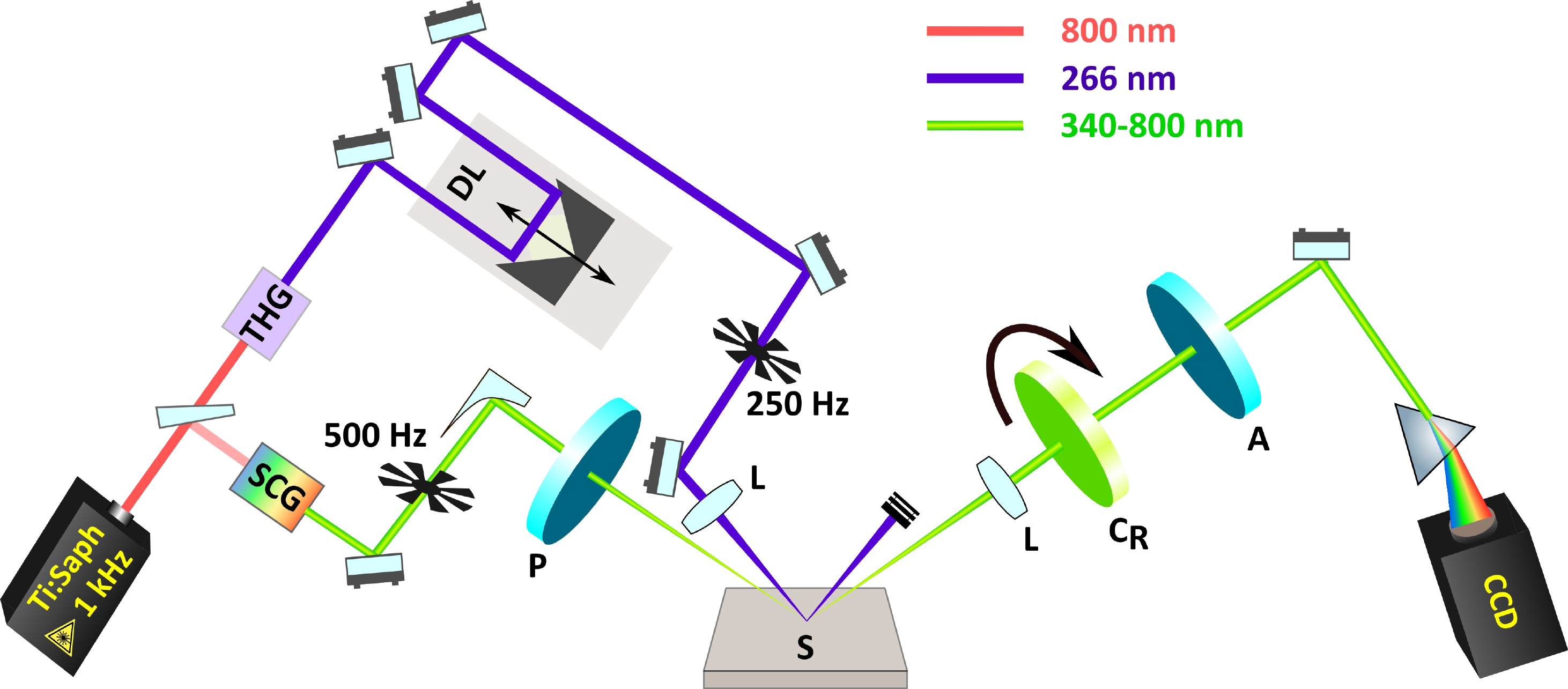}
\caption{Schematic of the experimental setup for the pump-probe time-resolved spectroscopic ellipsometry (trSE) measurements. THG: third harmonic generation for pump-beam, SCG: supercontinuum white-light generation for probe-beam, DL: delay-line, L: lense, P: polarizer, S: sample, C$_\mathrm{R}$: rotating compensator, A: analyzer.}
\label{fig_exp_setup}
\end{figure}

trSE measurements were performed using a femtosecond pulsed laser \cite{RSI_92_2021}, according to Fig. \ref{fig_exp_setup}. The third harmonic (THG, 266nm) of a titanium sapphire laser (Ti:Saph, Coherent Astrella, fundamental 35fs pulse duration, 800nm wavelength, 1kHz repetition rate) is utilized as the pump-beam, while 1$\%$ of the same laser pulse is used to generate supercontinuum white-light (SCG) in a CaF$_2$ window as the probe-beam. Using a prism spectrometer and a kHz-readout CCD camera, the transient reflectance-difference ($\Delta R/R$) spectra were recorded. A variable delay-line (DL) between the pump- and the probe-beam allows time-resolved measurements. Furthermore, a two-chopper system was employed for wavelength-dependent real-time correction of the pump-probe and only-probe measurements to obtain proper reflectance-difference spectra. For the purpose of acquiring the ellipsometric angles defined in Eq. (\ref{eq_ellip_theo}), the measured reflectance-difference spectra are applied to a reference steady-state spectra by a M$\mathrm{\ddot{u}}$ller-matrix formalism for each photon energy and delay-time. A more detailed description of the experimental setup and method can be found elsewhere \cite{APL_115_2019,NJP_22_2020,RSI_92_2021}. 

Ellipsometric angles $\Psi$ and $\Delta$ were measured in Polarizer-Sample-Compensator-Analyzer (PSCA) configuration (see Fig. \ref{fig_exp_setup}) for the probe-beam extending from 1.5-3.7eV. The pulses of 2.9\textmu J at 266nm (4.66eV) were used to excite the system inducing a carrier profile indicated schematically in Fig. \ref{fig_sample}. For a more detailed description, see Sec. \ref{sec_processes}. The angle of incidence was $60^{\circ}$ for the probe- and $55^{\circ}$ for the pump-beam. The 477\textmu m diameter of the pump-beam was determined by knife-edgde scan. Delay-times in the range of -10ps to 5000ps between pump und probe pulses were controlled by a motorized linear stage. The time resolution of the system was $\sim$100fs. Knowing $\Psi$ and $\Delta$, the pseudo dielectric function $\left\langle \varepsilon \right\rangle$ was calculated by:

\begin{equation}
\begin{split}
& \qquad \qquad \quad \rho=\frac{R_\mathrm{p}}{R_\mathrm{s}}=\tan\left(\Psi\right) \cdot\mathrm{e}^{\mathrm{i}\Delta} \\
& \left\langle \varepsilon \right\rangle = \sin^2\left(\Phi\right)\left(1+\tan^2\left(\Phi\right)\cdot\left(\frac{1-\rho}{1+\rho}\right)^2\right) . \\
\end{split}
\label{eq_ellip_theo}
\end{equation}

Here, $R_{\mathrm{p}}$ and $R_{\mathrm{s}}$ are the Fresnel reflection coefficients for p- and s-polarized light. $\left\langle \varepsilon \right\rangle$ contains information about the sample structure and therefore is not the DF of zb-GaN. The process for obtaining the DF from the measurement data is described in Sec. \ref{sec_evaluation}.


\section{Theory}
\label{sec_theory} 

The high-power pump beam induces a great number of electron-hole pairs in the material. These will have an effect on the transition energy between the conduction and the valence band due to many-body interactions.

\subsection{Carrier distribution}
\label{sec_processes}

The time-dependent free-carrier concentration is determined by different contributions concerning the absorption of photons by the material, the relaxation of pump-induced free-carriers in the respective bands and the recombination and diffusion of those free-carriers. Since the sample thickness of 605nm is much higher than the absorption depth of $\approx$60nm for the 266nm pump-beam, we have to consider a free-carrier profile within the sample. This leads to a more complicated analysis compared to homogeneous excited thin-film samples studied earlier \cite{APL_115_2019,NJP_22_2020,PRR_3_2021}. The starting point of our analysis to obtain a description for the free-carrier concentration $n\left(x,t\right)$ is the solution of the diffusion differential equation with the absorption profile as starting condition:

\begin{equation}
\begin{split}
& \quad \frac{\partial n\left(x,t\right)}{\partial t}=-\frac{n\left(x,t\right)}{\tau_1}+D\cdot\frac{\partial^2 n\left(x,t\right)}{\partial x^2}, \\
& \\
& \quad n\left(x=s,t=0\right)=\begin{cases}
I_0\cdot \mathrm{e}^{-\alpha s} & s \geq 0 \\
\ \ \ \ 0 & s < 0
\end{cases}. \\
\end{split}
\label{eq_dgl}
\end{equation}

Here, $x$ is the position in the GaN layer, $n$ means the pump-induced electron-hole concentration while $I_0$ resembles the number of absorbed photons at the surface of the sample. Furthermore, $s$ represents the absorption depth for the starting condition. This differential equation can be solved by Fourier-transformation for $\left(x,t\right)\geq 0$ to:

\begin{equation}
n_\mathrm{diff}\left(x,t\right)=\frac{1}{2}I_0\cdot \mathrm{e}^{\alpha^2Dt-\alpha x-\frac{t}{\tau_1}}\cdot \mathrm{erfc}\left(\frac{2\alpha Dt-x}{\sqrt{4Dt}}\right).
\label{eq_n_diff}
\end{equation} 

\begin{figure}[htb]
\includegraphics[angle=0, trim = 0mm 0mm 0mm 0mm, width=\columnwidth]{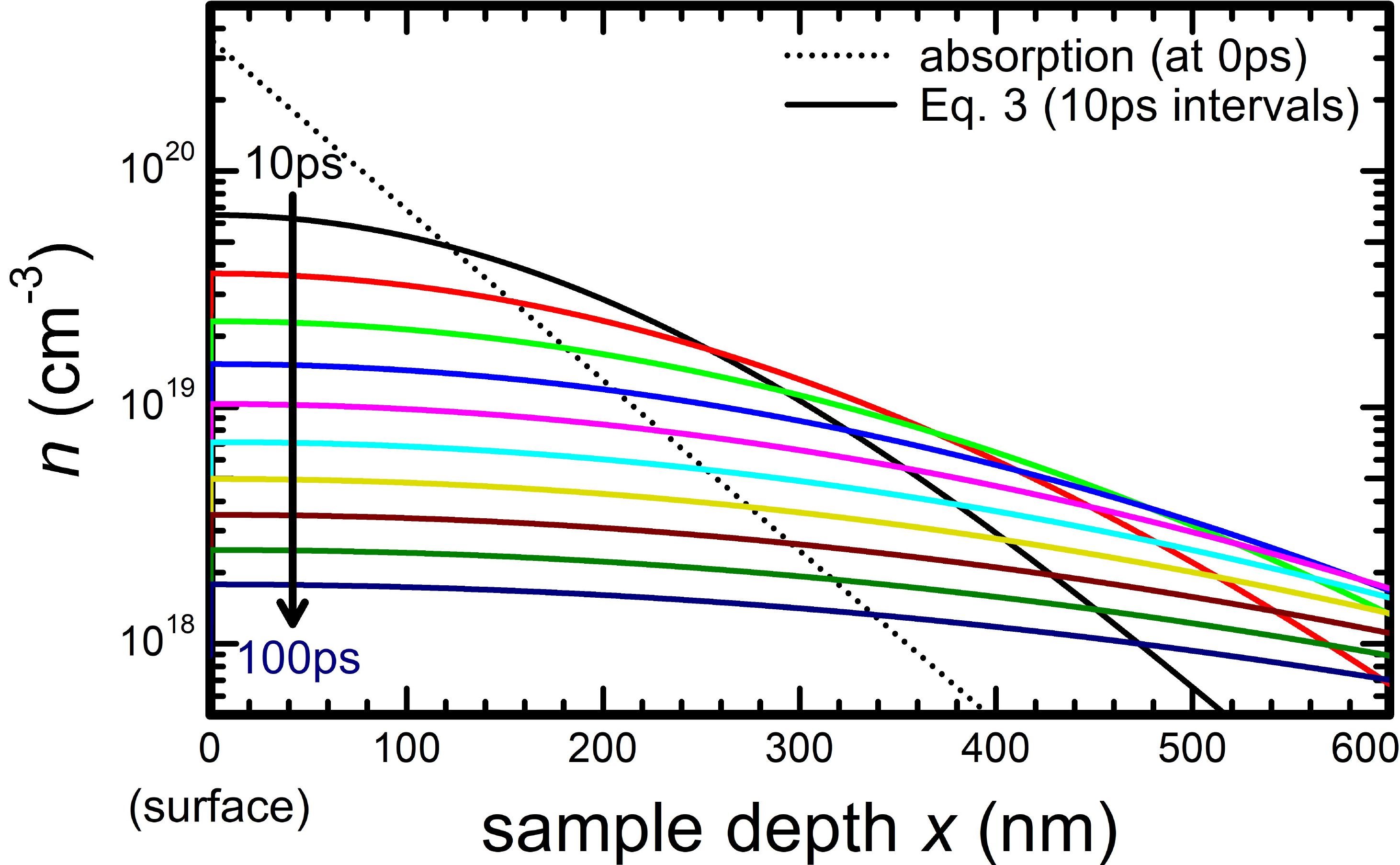}
\caption{Free-carrier distribution in the zb-GaN layer calculated by Eq. (\ref{eq_n_diff}) for different times, starting at 10ps (black) and ending at 100ps (dark blue) in steps of 10ps.}
\label{fig_Rel_Rek_Diff}
\end{figure}

In this solution, $\alpha=1/(60\mathrm{nm})$ is the absorption coefficient of zb-GaN at 4.66eV \cite{PRM_3_2019}, $D$ is the diffusion coefficient, and $\tau_1$ is the characteristic recombination time. The resulting free-carrier distribution dependent on the sample depth is shown in Fig. \ref{fig_Rel_Rek_Diff}. There, a strong time-dependent change in the free-carrier profile is visible, which allows us to determine an appropriate layer-model for GaN (see Sec. \ref{sec_evaluation}). We also assume, that free-carriers are reflected back into the sample at the sample surface. This condition provides a free-carrier maximum at $x=0$ and is equal to negligible surface recombination in zb-GaN \cite{PSS_10_2016}.

Obviously, Eq. (\ref{eq_n_diff}) does not describe the relaxation effect. To account for it, we consider only the pump-induced electrons in the conduction band, since the Fermi-vector of electrons is always greater than the respective Fermi-vector of holes due to the fact, that the total amount of pump-induced holes is distributed over three valence bands. This means, that the electron distribution in the conduction band dictates the absorption process of our probe-beam and therefore the measured transition energy. \\
The total amount of pump-induced electrons $N_0$ relax exponentially into the empty conduction band minimum (CBM) having a characteristic relaxation time $\tau_0$. Consequently, the number of electrons in the CBM can be expressed by $N_0\cdot \left(1-\mathrm{e}^{-t/\tau_0}\right)$. However, an accurate model for the relaxation has to consider the pump-beam profile since relaxation time and pump-beam duration are both in the same order of magnitude. The relaxation model follows from the convolution of the number of electrons in the CBM and the pump-beam profile. Assuming the pump-beam profile to be Gaussian, the resulting relaxation model strongly resembles an error-function. Therefore, we approximate the number of relaxed electrons by:

\begin{equation}
n_\mathrm{relax}\left(t\right)=\frac{N_0}{2}\cdot \left(\mathrm{erf}\left(\frac{t-\gamma_0}{\tau_0}\right)+1\right)
\label{eq_n_relax}
\end{equation} 

with $N_0$ and $\tau_0$ as defined before and $\gamma_0$ as the position of the inflection point of the error function. We obtain the model for the free-carrier concentration in the CBM, which accounts for the change in the absorption onset, at the sample surface ($x=0$) by replacing $I_0$ in Eq. (\ref{eq_n_diff}) by the relaxation model: 

\begin{equation}
n_\mathrm{model}\left(t\right)=\frac{n_\mathrm{relax}\left(t\right)}{2}\cdot \mathrm{e}^{\alpha^2Dt-\frac{t}{\tau_1}}\cdot \mathrm{erfc}\left(\sqrt{\alpha^2Dt}\right).
\label{eq_model_n}
\end{equation} 

\begin{figure}[htb]
\includegraphics[angle=0, trim = 0mm 0mm 0mm 0mm, width=\columnwidth]{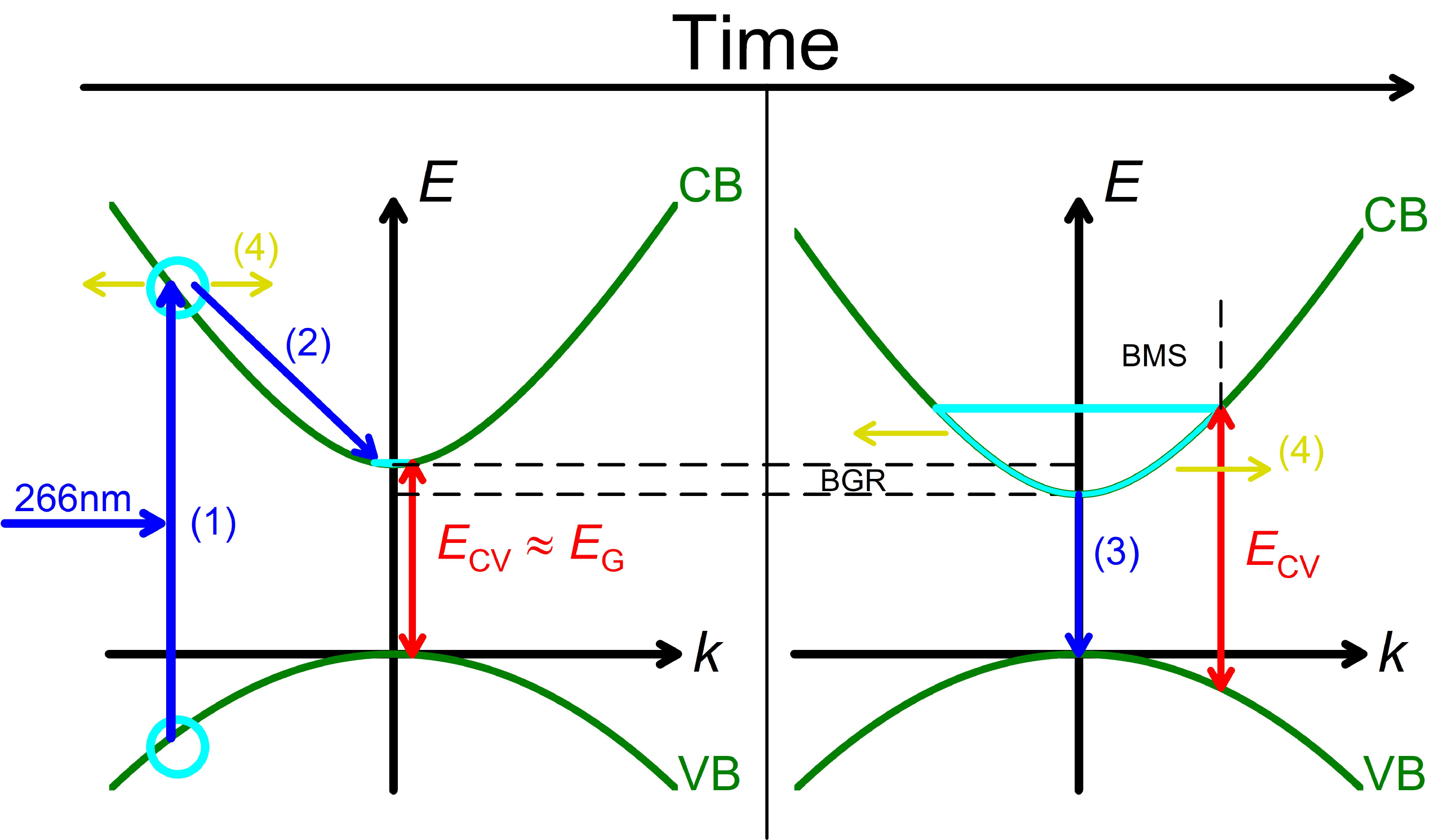}
\caption{Involved processes (only shown for electrons) influencing the transition energy between the conduction band (CB) and the valence bands (VB), here portrait by a single band: absorption (1) of the 266nm pump-beam, relaxation (2) into the CB minimum, recombination (3) back into the VB and diffusion (4) within the sample (no $k$-variation). Many-body effects shift possible transitions depending on the number of electrons in the CB-minimum.}
\label{fig_band_diagramm}
\end{figure}

A schematic overview of the involved processes is given in Fig. \ref{fig_band_diagramm}.

\subsection{Many-body effects}
\label{sec_band}

The band structure of zb-GaN in the vicinity of the $\Gamma-$point of the BZ is known to consist of one conduction band and three valence bands. Setting the direct band gap to the zone center, assuming isotropic bands and a scalar effective mass, we can describe the conduction band dispersion by \cite{JPCS_1_1957,PRB_66_2002_201403}:

\begin{equation}
E_\mathrm{C}(k)=\frac{\hbar^2k^2}{2m_\mathrm{e}}+\frac{1}{2}\left(E_\mathrm{G}+\sqrt{E_\mathrm{G}^2+4P^2k^2}\right).
\label{eq_ECB}
\end{equation}

Here, $m_\mathrm{e}$ is the (free-) electron rest mass and $P$ is a momentum matrix element, which is assumed to be $k$-independent \cite{yu_cardona}. This $\textbf{k} \cdot \textbf{p}$ perturbation theory based model already contains the nonparabolicity of the conduction band. \\
The influence of many-body interactions on the interband transition energies is a key factor for understanding the experimental data as well as the development of possible applications. Due to phase-space filling of the conduction band, the Fermi-energy and -vector are increasing with increasing free-carrier concentration and thus increase the interband transition energy \cite{PPSB_67_1954,PR_93_1954}. This so-called Burstein-Moss shift (BMS) contains a conduction band and a valence band contribution. In a n-type doped semiconductor, the BMS is described by the following expressions:

\begin{equation}
E_{\mathrm{BMS}}\left(n\right)=E_\mathrm{C}(k_\mathrm{F})+\left|\bar{E_\mathrm{V}}(k_\mathrm{F})\right|.
\label{eq_bms}
\end{equation}

The conduction band contribution is given in Eq. (\ref{eq_ECB}) while an averaged parabolic valence band contribution over the light hole, heavy hole, and split-off band is sufficient to describe the more complicated valence band contribution \cite{PRM_3_2019}. \\
On the other hand, the band gap renormalization (BGR) \cite{PRB_24_1981,PRB_66_2002_201403} in a n-type doped semiconductor can be written as \cite{PRB_93_2016,AP_27_1978}:

\begin{equation}
E_{\mathrm{ren}}\left(n\right)=E_\mathrm{G}-\Delta E_{\mathrm{BGR}}(n)=E_\mathrm{G}-\Delta E_{\mathrm{ee}}(n)-\Delta E_{\mathrm{ei}}(n).
\label{eq_bgr}
\end{equation} 

Here, electron-electron ($\Delta E_{\mathrm{ee}}$) and electron-ion ($\Delta E_{\mathrm{ei}}$) interactions cause a decrease ($-\Delta E_{\mathrm{BGR}}$) of the fundamental band gap $E_\mathrm{G}$. The equations for those contributions can be found elsewhere \cite{PRM_3_2019}. \\
However, this study does not investigate n-type doped semiconductors, but pump-induced electron and hole accumulations in the CBM or valence band maximum, respectively. Therefore, we do not expect an electron-ion interaction as contained in Eq. (\ref{eq_bgr}) but suppose additional hole-hole and electron-hole interactions. It also is reasonable, that different valence band contributions yield different BGR effects due to different hole masses and different distributions of holes among the three valence bands. This, of course, would also dramatically change an attempt of averaging over all valence bands. Unfortunately, the inclusion and accurate description of all these effects would far exceed the framework of this study. We therefore make some reasonable approximations to obtain an adequate model for the many-body interactions. First-principle calculations suggest that hole-hole interactions yield only a fraction of the effect of electron-electron interactions \cite{JAP_92_2002}. Assuming that in combination with the fact, that each valence band is less filled with holes than the conduction band is filled with electrons, leads us to estimate the combined hole-hole and electron-hole interaction as the electron-ion interaction in Eq. (\ref{eq_bgr}). This approximation should be good enough for low free-carrier concentrations at which the BGR is dominant over the BMS. For higher free-carrier concentrations the difference in BGR calculation should be negligible compared to the BMS contribution. We therefore assume the transition energy $E_{\mathrm{CV}}\left(n\right)$ as a function of the free-electron concentration in the CBM as:

\begin{equation}
E_{\mathrm{CV}}\left(n\right)=\frac{1}{2}\left(\frac{\hbar^2k_{\mathrm{F}}^2}{m_\mathrm{e}}+\frac{\hbar^2k_{\mathrm{F}}^2}{\bar{m}_{\mathrm{h}}}+E_{\mathrm{ren}}+\sqrt{E_{\mathrm{ren}}^2+4P^2k_{\mathrm{F}}^2}\right)
\label{eq_ecv}
\end{equation}

with the Fermi-vector $k_\mathrm{F}\left(n\right)$=$\left(3\pi^2n\right)^{\frac{1}{3}}$, an averaged hole mass of $\bar{m}_{\mathrm{h}}=0.61m_\mathrm{e}$ and $P=0.724$eVnm \cite{PRM_3_2019}.


\section{Analysis of experimental optical data}
\label{sec_evaluation} 

\begin{figure}[htb]
\includegraphics[angle=0, trim = 0mm 0mm 0mm 0mm, width=\columnwidth]{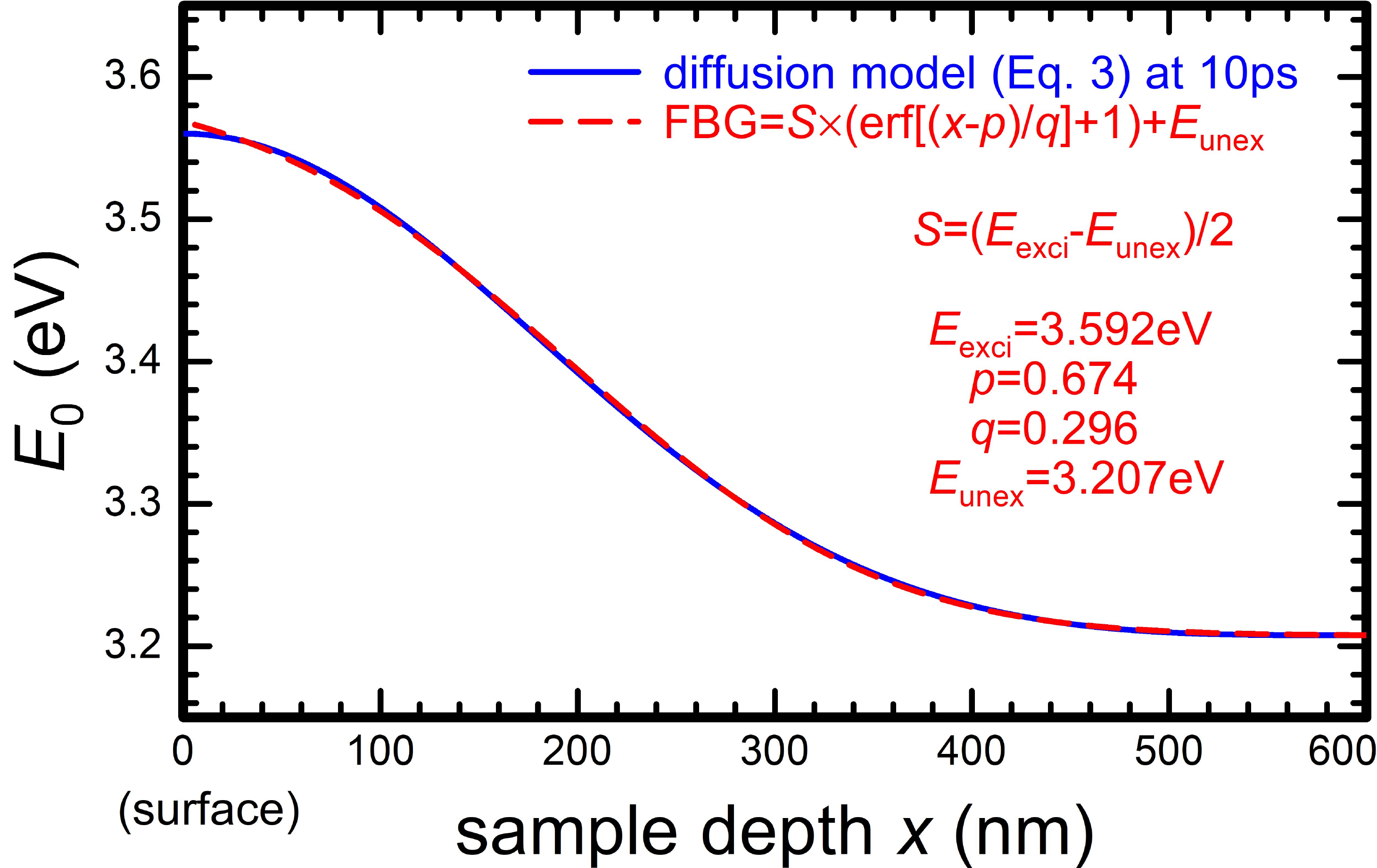}
\caption{Determination of the $p$ and $q$ parameters for the function-based graded curve (FBG) at 10ps. The transition energies for the excited ($E_\mathrm{exci}$) surface and the less or non-excited ($E_\mathrm{unex}$) bottom are fitted as well.}
\label{fig_grading_fbg}
\end{figure}

This section describes the steps to investigate the experimental data and to obtain the transition energy and therefore the free-carrier concentration at any delay-time. The measured ellipsometric angles $\Psi$ and $\Delta$ are analyzed by constructing a multi-layer model, representing the layout of the sample consisting of zb-GaN, 3C-SiC, and Si from surface to substrate, by use of the Woollam WVASE32 software. Additionally, a Bruggeman effective medium approximation (EMA) layer including 50\% void in the GaN-matrix on top of the sample is added to account for surface roughness \cite{AP_24_1935}. The zb-GaN layer is characterized by a general oscillator model (GenOsc) to fit the measured data. The oscillator used to describe the band gap region was a Psemi-M0 oscillator \cite{WVASE}. The main parameters in this are $A_0$, $E_0$ and $B_0$, which account for the amplitude, energy position and broadening of the supposed band edge. The applied model was fitted to the experimental data by numerically minimizing mean squared error values using a Levenberg-Marquardt algorithm. The resulting model DF was used as a starting point to perform a point-by-point (pbp) fit, which numerically changes the value of the DF wavelength-by-wavelength until the best agreement with the experimental data is achieved \cite{JAP_114_2013,PRB_90_2014_195306}. Then, the resulting pbp-DF is again line shape-fitted by the GenOsc to obtain the transition energy. \\
However, since the pump-beam creates an excited carrier profile in the sample, it is necessary to implement the same profile in the GenOsc model of the zb-GaN layer. For this, we use the so-called function-based graded layer (FBG) \cite{WVASE}. The FBG allows us to vary the GenOsc parameters dependent on the position in the layer. In this case $A_0$, $E_0$, and $B_0$ are varied as a function of layer thickness, between the excited GaN at the surface and the less or non-excited GaN at the bottom (see Fig. \ref{fig_sample}). With this, we are able to translate the diffusion model from Eq. (\ref{eq_model_n}) into the WVASE software. Of course, the free-carrier concentration has to be converted to the transition energy by means of BMS and BGR first. We also approximate the diffusion model by an error-function dependent on the transition energy of the excited and the not excited GaN, as well as the broadening parameter $q$ and the inflection point position $p$ as shown in Fig. \ref{fig_grading_fbg}.

We then fit the layer model containing the FBG (including the $p,q-$parameters obtained from the diffusion model) to the experimental data ($\Psi$ and $\Delta$) by varying the parameters of the excited GaN. The shape of the depth profile is kept constant, only the surface ($x=0$) values are varied. This is the starting point for further analysis. Unfortunately, within the FBG layer we are not able to perform a pbp-fit which is required to accurately determine the transition energy. Therefore, we replace the FBG layer after the first fit by an EMA-graded layer. Here, the grading is performed by varying the content percentage of excited GaN in an unexcited GaN-matrix over the sample depth, with 100\% excited GaN at the top and 0\% at the bottom. The gradient is also modelled by an error-function. This form of grading does not represent the actual absorption onset profile but provides a sufficient approximation of the experimental data. We stress the fact that already determined values for amplitude, transition energy, and broadening factor of the excited GaN are kept constant. Before starting the pbp-fit, the EMA-graded model is fitted to the measurement data one last time, allowing only the grading-parameters to be varied. This should account for the unavoidable difference between the FBG- and the EMA-approach. The pbp-result for the excited GaN is displayed and described in the next section. Keep in mind, that different sample depths yield different DFs. In this study, only the DF at $x=0$ (surface) is analyzed. Furthermore, we try to simplify the complicated FBG-approach by easier approximations for certain time regions. For delay-times $t>100\mathrm{ps}$, the diffusion of the pump-induced carriers yields an almost homogeneous distribution in the GaN layer, as shown in Fig. \ref{fig_Rel_Rek_Diff}. Therefore, it is sufficient to use a single or double layer approach for the GaN layer. On the other hand, at the very beginning after the pump-beam excitation $\left(t<10\mathrm{ps}\right)$, diffusion and recombination are negligible. The free-carrier profile is mostly determined by the absorption profile. Therefore, the FBG does not contain the diffusion model but a simple absorption profile based on exponential decline.


\section{Results and Discussion}
\label{sec_results} 

\begin{figure}[htb]
\includegraphics[angle=0, trim = 0mm 0mm 0mm 0mm, width=\columnwidth]{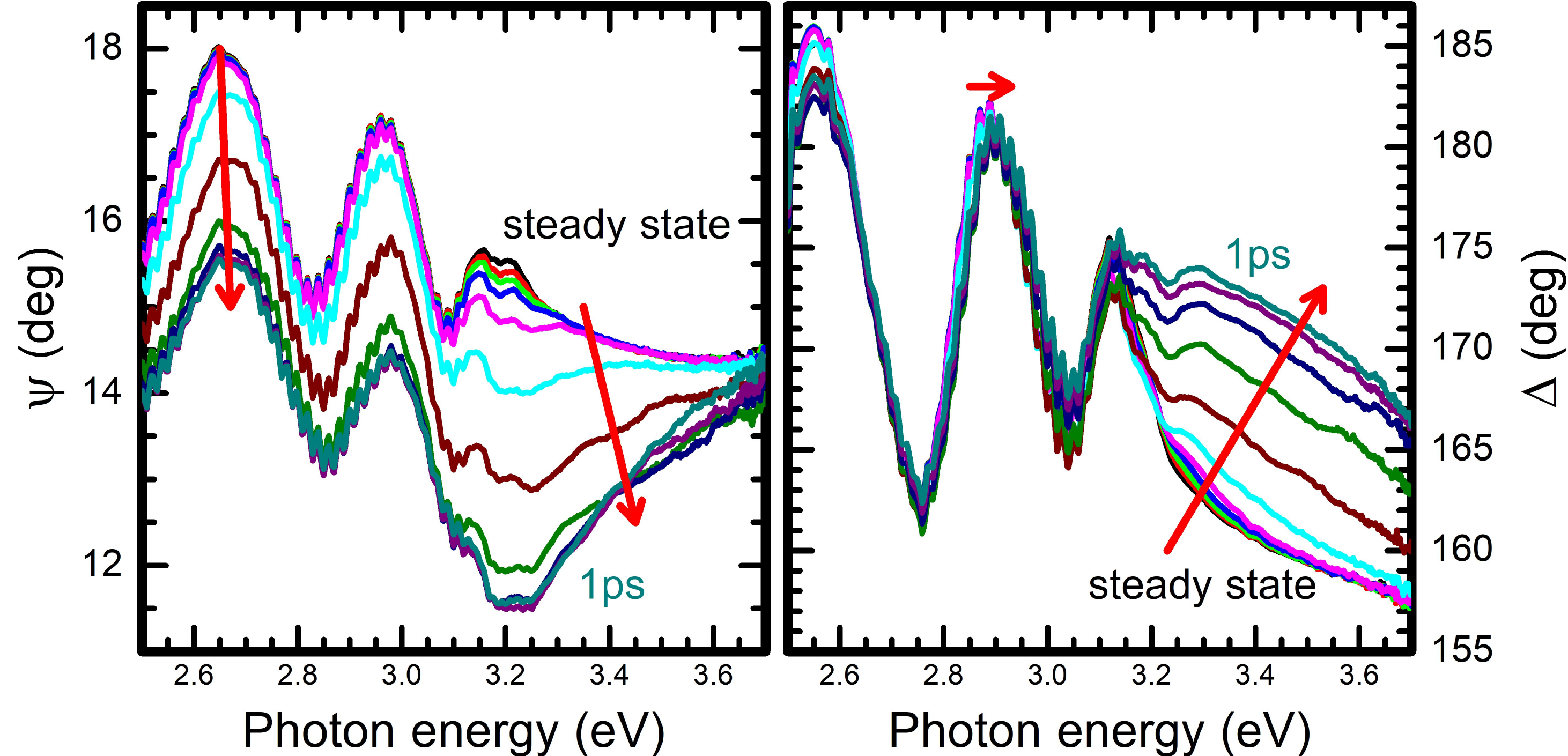}
\caption{Ellipsometric angles $\Psi$ (left) and $\Delta$ (right) in 0.1ps time steps for the steady-state (black) to 1ps (teal) measurements. Arrows indicate the spectral evolution.}
\label{fig_psi_delta_very_short}
\end{figure}

In this section, we explain the analysis of the experimental data and resulting DFs for selected delay-times between the incident pump pulses and the probe pulses that follow the return of the excited system to its ground (non-excited) state. We also provide here our interpretation of the transition energies. In the beginning, we assume that the measurements performed at a negative delay (-10ps) are equivalent of data obtained by steady-state experiments on the non-excited system. The experimental data ($\Psi$ and $\Delta$) for the first picosecond after the excitation are shown in Fig. \ref{fig_psi_delta_very_short}. Here, all spectra display Fabry-P\'{e}rot oscillations below 3.2eV and especially the relaxation effect (see Sec. \ref{sec_processes}) dominates the spectral evolution in this timescale. The first signature of a dynamic relaxation process is the increasing difference between excited and steady-state spectra that maximizes at about 1ps. After 1ps, $\Psi$ and $\Delta$ start returning to their steady-state values. 

\begin{figure}[htb]
\includegraphics[angle=0, trim = 0mm 0mm 0mm 0mm, width=\columnwidth]{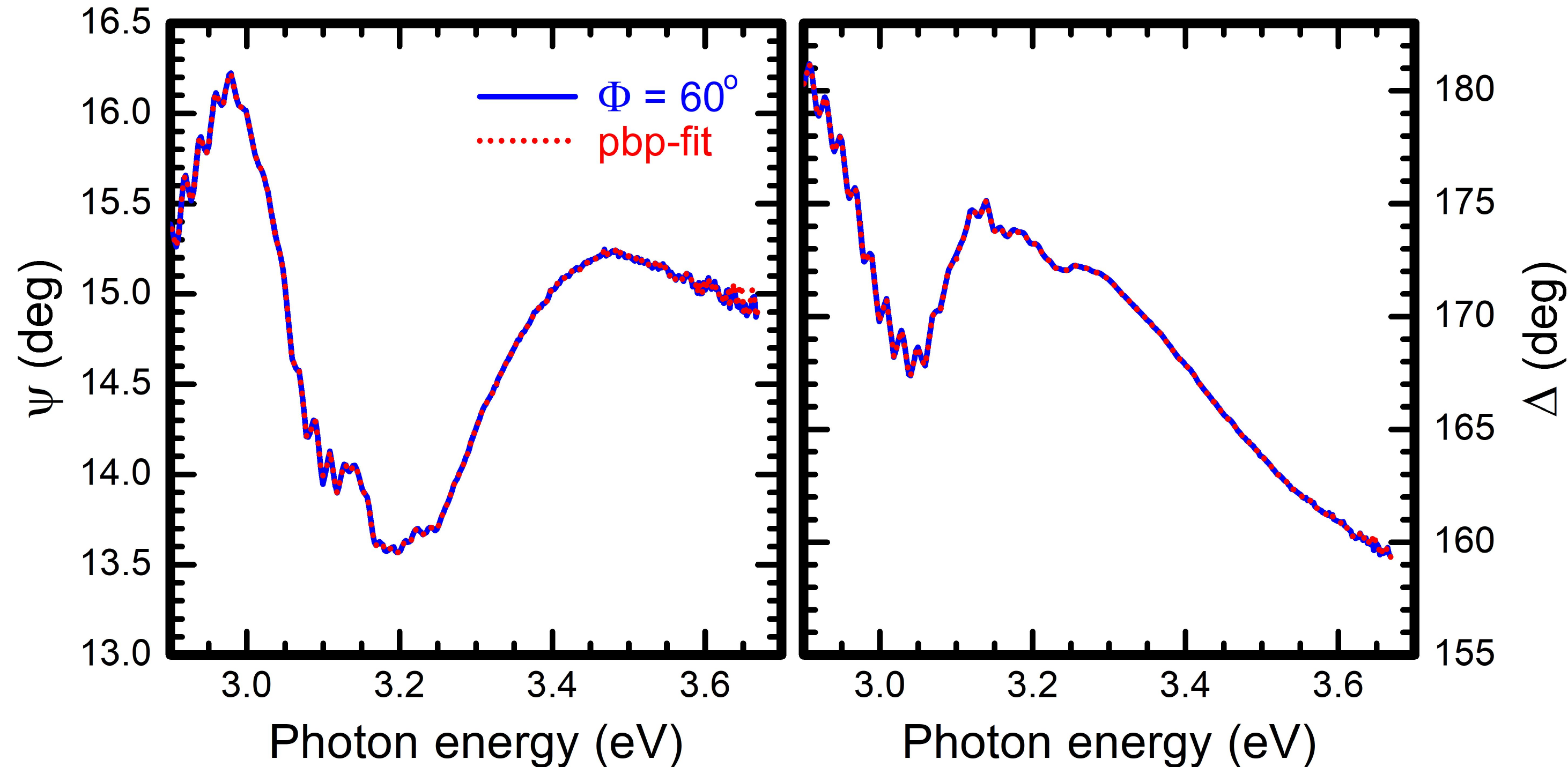}
\caption{Measurement data (blue, continuous) and point-by-point-fit (red, dotted) in $\Psi$ and $\Delta$ for the 10ps measurement.}
\label{fig_10ps_psi_delta}
\end{figure}

\begin{figure}[htb]
\includegraphics[angle=0, trim = 0mm 0mm 0mm 0mm, width=\columnwidth]{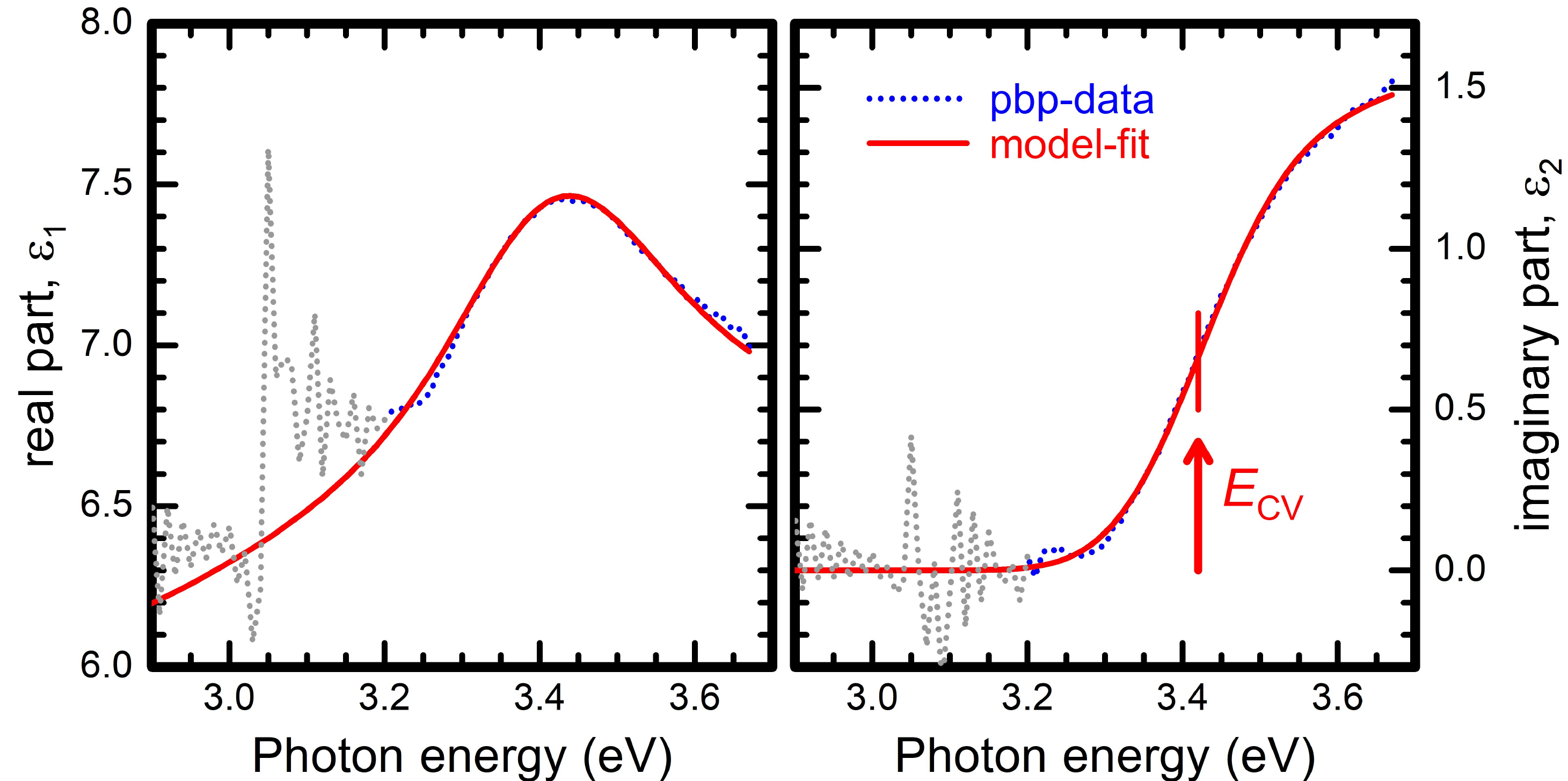}
\caption{Point-by-point fitted (dotted) and analytical model (continuous) dielectric function for the 10ps measurement. The transition energy $E_\mathrm{CV}$ is indicated by an arrow.}
\label{fig_10ps_eps1_eps2}
\end{figure}

As an example, we present the analysis of the 10ps measurement in Figs. \ref{fig_10ps_psi_delta} and \ref{fig_10ps_eps1_eps2}. Here, the model described in Sec. \ref{sec_evaluation} is applied on the experimental data ($\Psi$ and $\Delta$) with the resulting pbp-fit. We focus the analysis on the spectral range near the band gap (2.9-3.7eV) to achieve the most accurate pbp-fit to determine the transition energy. 

Although a very convincing pbp-fit was achieved (Fig. \ref{fig_10ps_psi_delta}), the DF displays inconsistencies below 3.2eV, especially in $\varepsilon_1$ (Fig. \ref{fig_10ps_eps1_eps2}). This is most probably due to shifted starting points of the Fabry-P\'{e}rot oscillations caused by inconsistent transparencies over the sample depth, which are not accounted for in our model. Our main result is the absorption onset position, thus we are not focusing on the Fabry-P\'{e}rot resonances further. The position of the transition energy is approximated to be the inflection point in $\varepsilon_2$, which coincides with the $E_0$-parameter of the used Psemi-M0 oscillator.

\begin{figure}[htb]
\includegraphics[angle=0, trim = 0mm 0mm 0mm 0mm, width=\columnwidth]{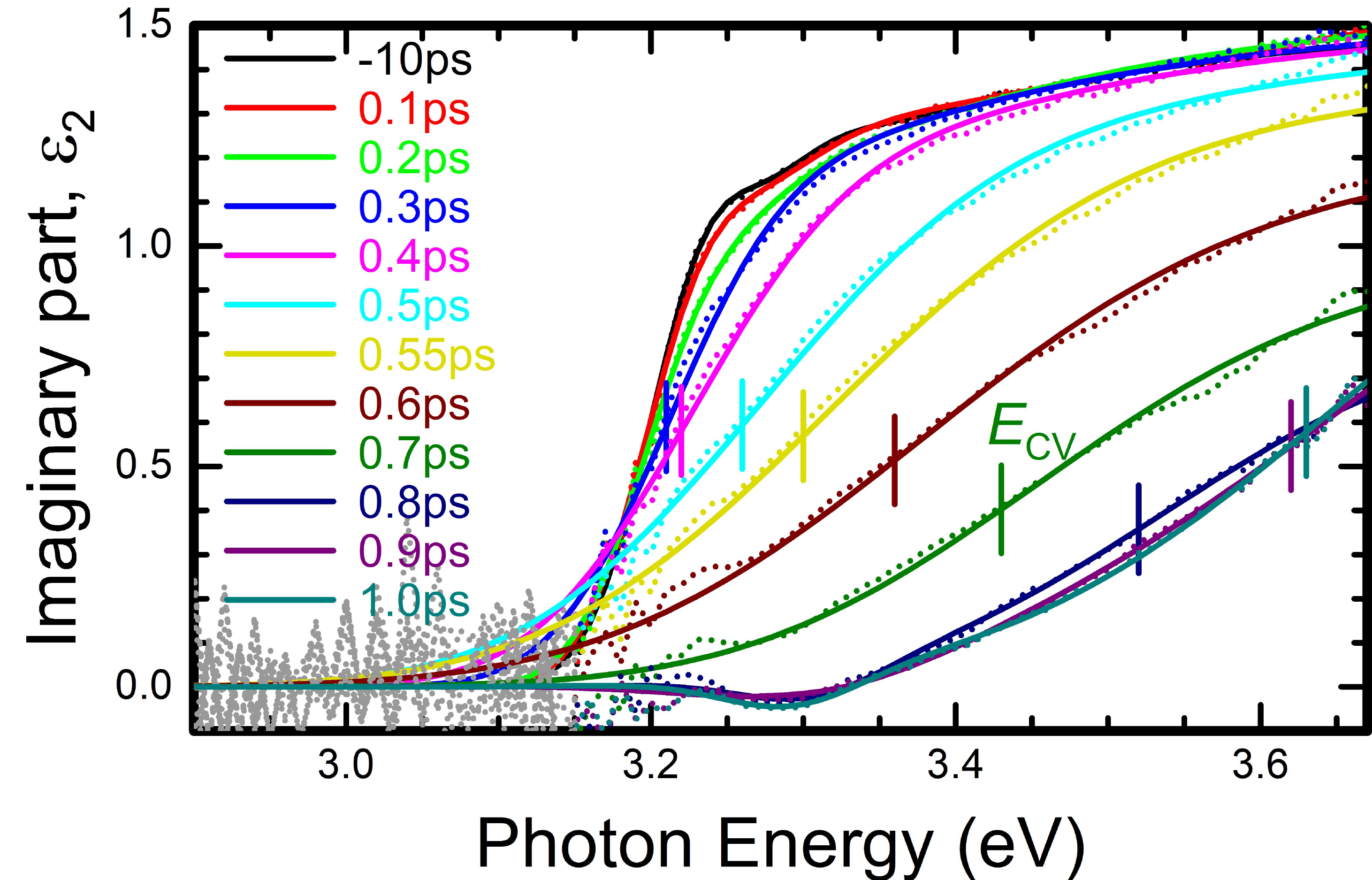}
\caption{Point-by-point fitted (dotted) and analytical model (continuous) imaginary parts of the dielectric functions in the first picosecond after the pump pulse. The transition energies $E_\mathrm{CV}$ are indicated by vertical lines for selected delay-times.}
\label{fig_DF_very_short}
\end{figure}

The same procedure is performed for every delay-time shown in this study. The resulting imaginary parts of the pbp-DFs and their GenOsc model-fits are shown in Figs. \ref{fig_DF_very_short}-\ref{fig_DF_long} for three different time windows. The last shown curve is always the first one of the next figure with exception of the steady-state measurement (-10ps), which is present in all figures for better comparison. The pbp-fit result below 3.2eV will be ignored as discussed above. The FBG-layer contains the simple absorption model for $t<1.2$ps. At that point the diffusion model described in Sec. \ref{sec_processes} is used up to 39ps. From there the FBG-layer is replaced by a double-layer approach which transitions into a single layer at $t=1000$ps.

For $t<1$ps (Fig. \ref{fig_DF_very_short}), we can clearly see the shift of the absorption onset to higher energies as a function of time as well as the increased broadening of the absorption edge due to the increasing carrier-temperature. Effects of these hot charge carriers have been reported earlier for oxide semiconductors \cite{NJP_22_2020}. The large broadening of $\varepsilon_2$ in combination with the upper measurement range limit of 3.7eV are the reasons why only an uncertain determination of the transition energy for the spectra with the highest energy inflection points is achievable. It is possible, that these transition energies are too low compared to the actual value. The change in $\varepsilon_2$ from 0.5ps to 1ps is much stronger than the change from 0.1ps to 0.5ps. This is true for amplitude, energy position, and broadening and follows the relaxation process due to the pump-beam profile as described in Sec. \ref{sec_processes}. It should be noted, that there is an additional signal in the steady-state measurement at $\approx$3.3eV. We attribute this to the combination of exciton and band-band transition by describing it by an additional Gaussian oscillator. Thus, the signal at $\approx$3.3eV resembles the high energy shoulder of the band-band transition. The resulting exciton binding energy is $\approx$15meV. This excitonic structure disappears completely after the 0.2ps measurement due to dominant free-carrier and band-filling effects. 

\begin{figure}[htb]
\includegraphics[angle=0, trim = 0mm 0mm 0mm 0mm, width=\columnwidth]{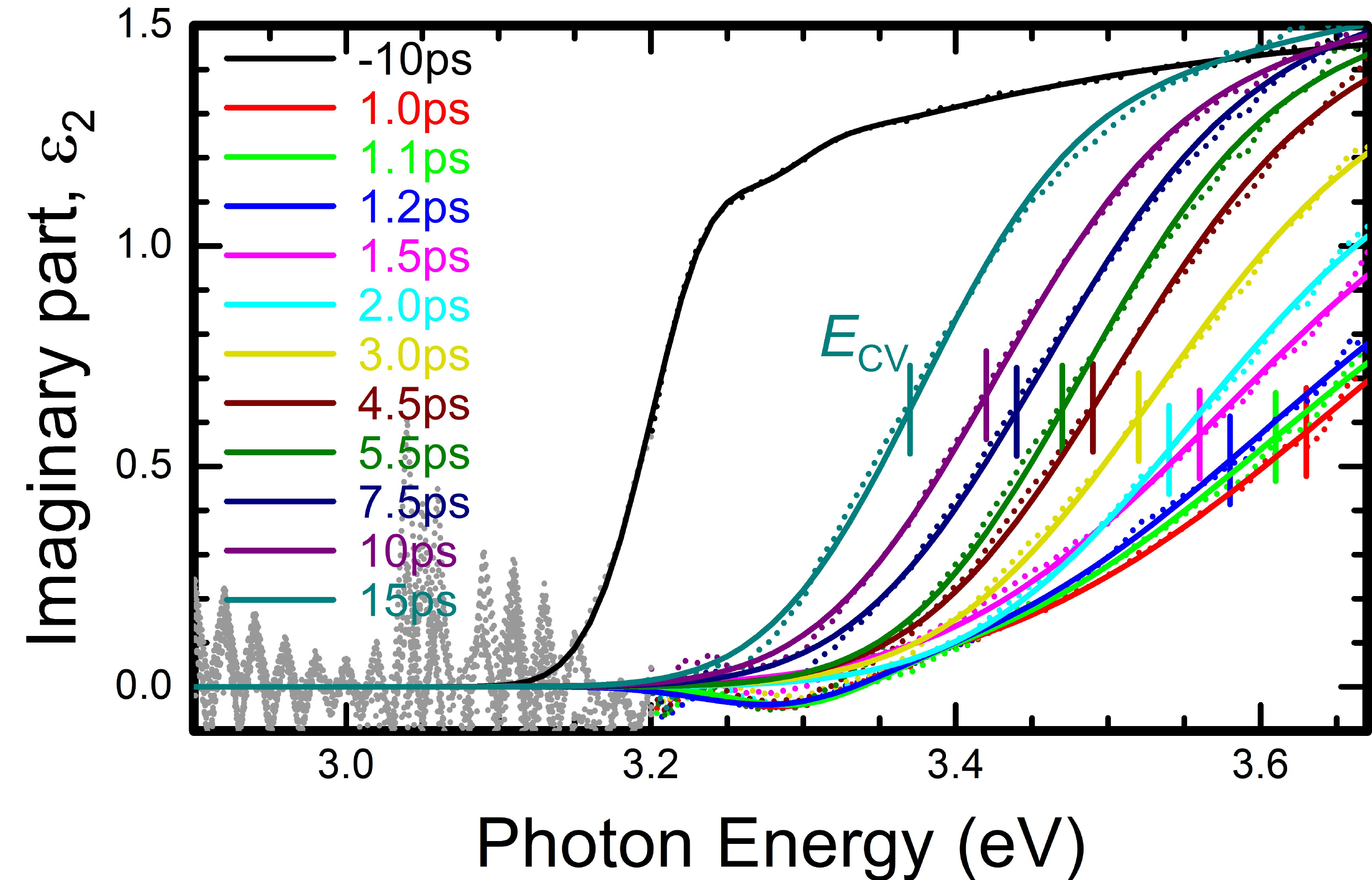}
\caption{Point-by-point fitted (dotted) and analytical model (continuous) imaginary parts of the dielectric functions up to 15 picoseconds after the pump pulse.The transition energies $E_\mathrm{CV}$ are indicated by vertical lines.}
\label{fig_DF_short}
\end{figure}

A closer inspection of the DF at 0.8-1.2ps yields a negative contribution in $\varepsilon_2$ in both the pbp-data and the model-fit. We attribute this to the occurrence of material gain and not to inconsistencies in the fit. Reasons for this assumption are on the one hand the relatively smooth pbp-result and on the other hand, the necessity to include the gain in the line shape-fit to obtain systematic transition energies. If gain is not accounted for in the line shape-fit, the used GenOsc for describing the band edge displays an erratic behaviour regarding the amplitude, energy position, and broadening. To include the gain effect in the model, we approximate it by a Gaussian oscillator having negative amplitude. The position of the Gaussian is relatively stable for all five time steps ($0.8\leq t\leq 1.2$) at roughly 3.3eV. 

\begin{figure}[htb]
\includegraphics[angle=0, trim = 0mm 0mm 0mm 0mm, width=\columnwidth]{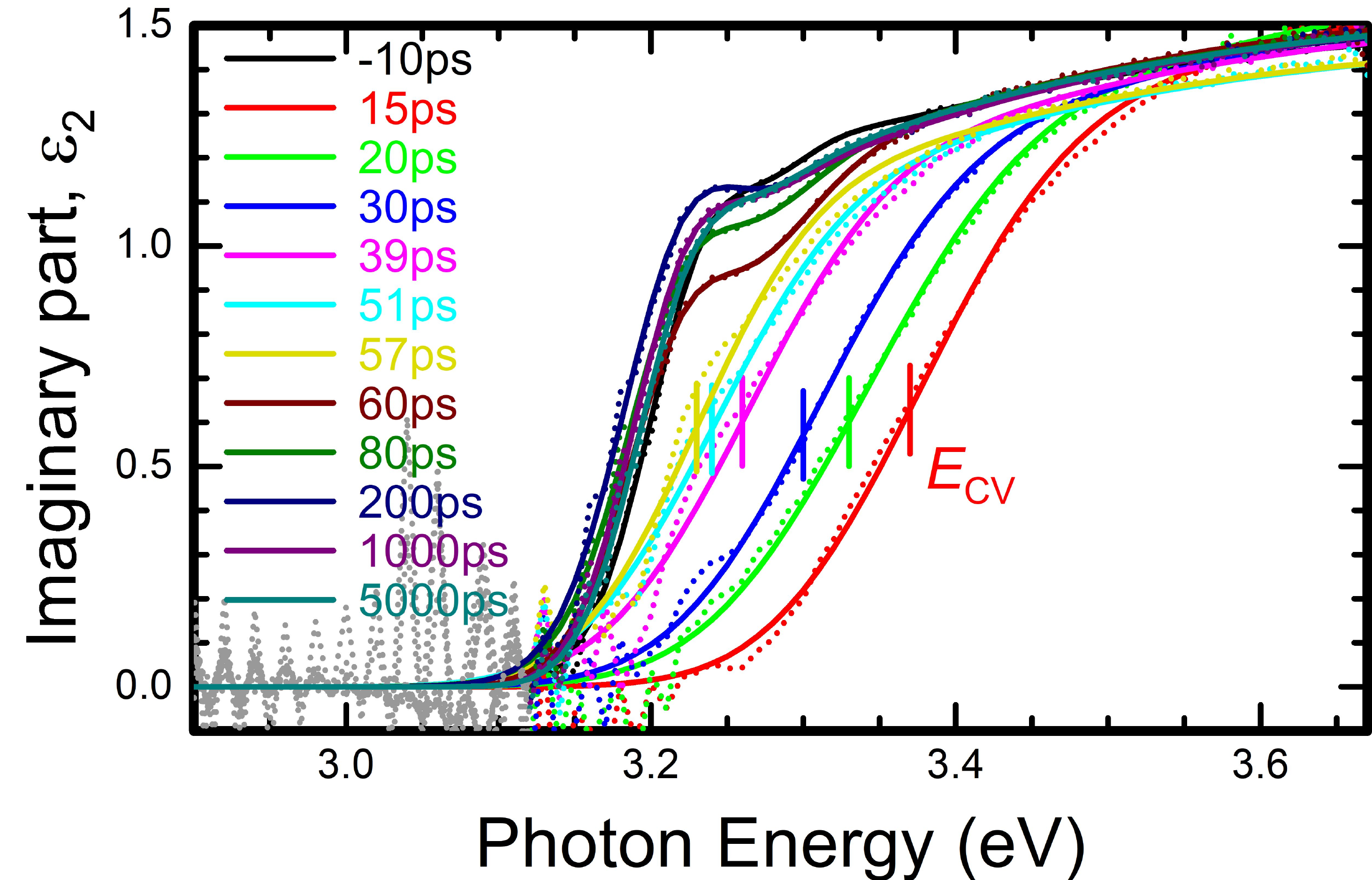}
\caption{Point-by-point fitted (dotted) and analytical model (continuous) imaginary parts of the dielectric functions long after the pump pulse. The transition energies $E_\mathrm{CV}$ are indicated by vertical lines for selected delay-times.}
\label{fig_DF_long}
\end{figure}

After the 5.5ps measurement, the broadening of the absorption edge seems to have returned to the value of the steady-state case. From this point on, we observe almost a parallel shift of the absorption edge to lower energies in $\varepsilon_2$. The amplitude also slowly returns to the steady-state value while the transition energy decreases due to diffusion and recombination. At $t=60$ps, the excitonic signal from the steady-state measurement appears in the pbp-data and has to be taken into account for the model by an additional Gaussian oscillator. Here, the transition energy reached the same position as in the ground state. For longer delays, changes in the DF happen relatively slowly. The exciton contribution becomes more dominant and the transition energy decreases further, even slightly below the steady-state value. This is possible for a specific free-carrier concentration where the BGR contribution is stronger than the BMS one. Finally, the induced free-carriers are completely recombined.

\begin{figure}[htb]
\includegraphics[angle=0, trim = 0mm 0mm 0mm 0mm, width=\columnwidth]{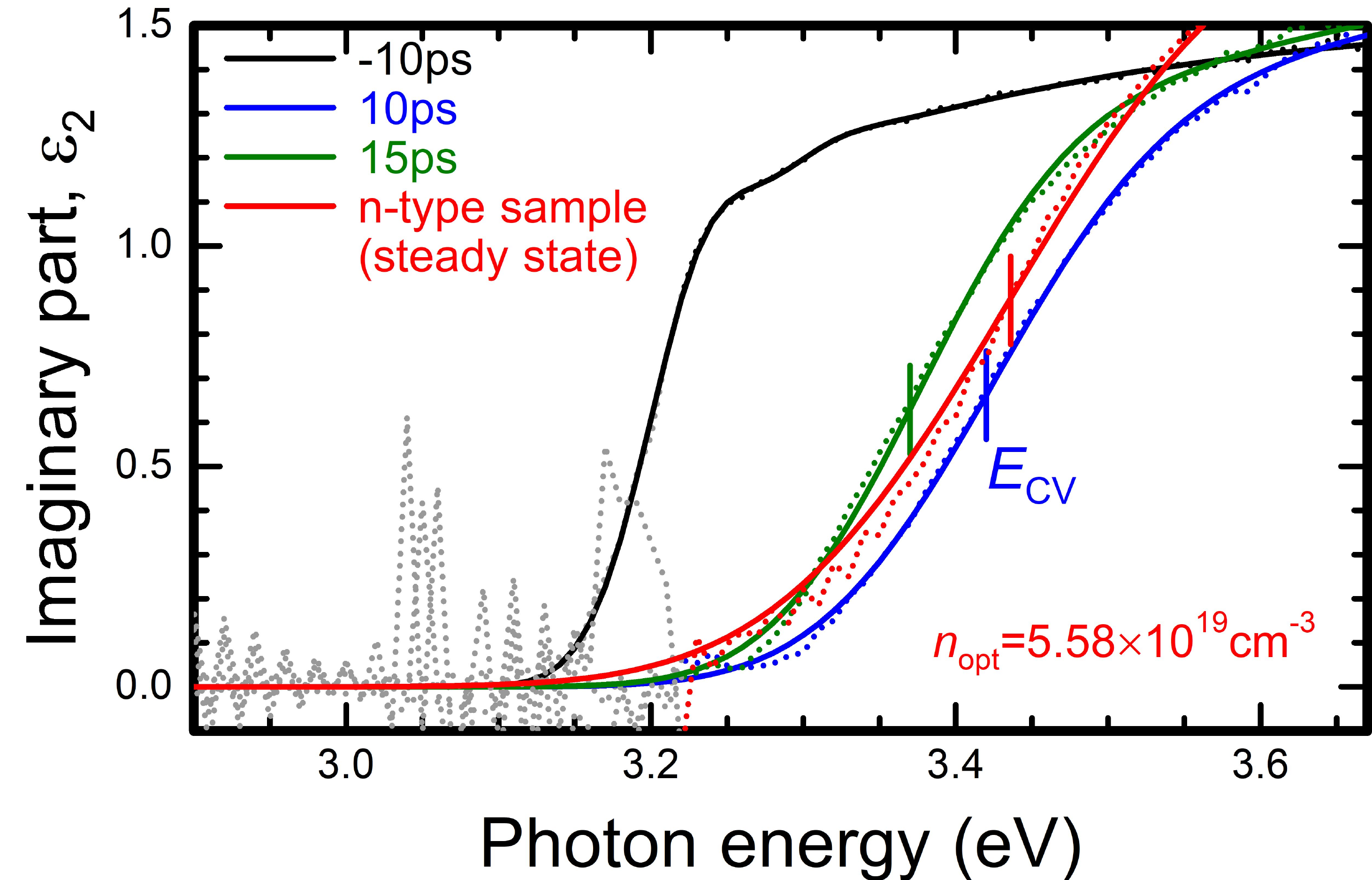}
\caption{Comparison of pbp- (dotted) and model- (continuous) imaginary parts of the dielectric functions of time-resolved measurement data to a steady-state measurement performed on a n-type zb-GaN sample with an optical free-carrier concentration of $5.58\times 10^{19}$cm$^{-3}$ \cite{PRM_3_2019}.}
\label{fig_comp_2523}
\end{figure}

We compare the DFs of $t=10$ps and $t=15$ps to the DF of a highly n-type doped zb-GaN sample (Fig. \ref{fig_comp_2523}). The n-type sample shown here was discussed in a previous study \cite{PRM_3_2019}. Since both the transition energy and the broadening of these DFs are similiar to each other, we can estimate the free-carrier concentration to be in the order of $5\times 10^{19}$cm$^{-3}$ for further analysis.

\subsection{Carrier influenced transition energies}
\label{sec_gain}

\begin{figure}[htb]
\includegraphics[angle=0, trim = 0mm 0mm 0mm 0mm, width=\columnwidth]{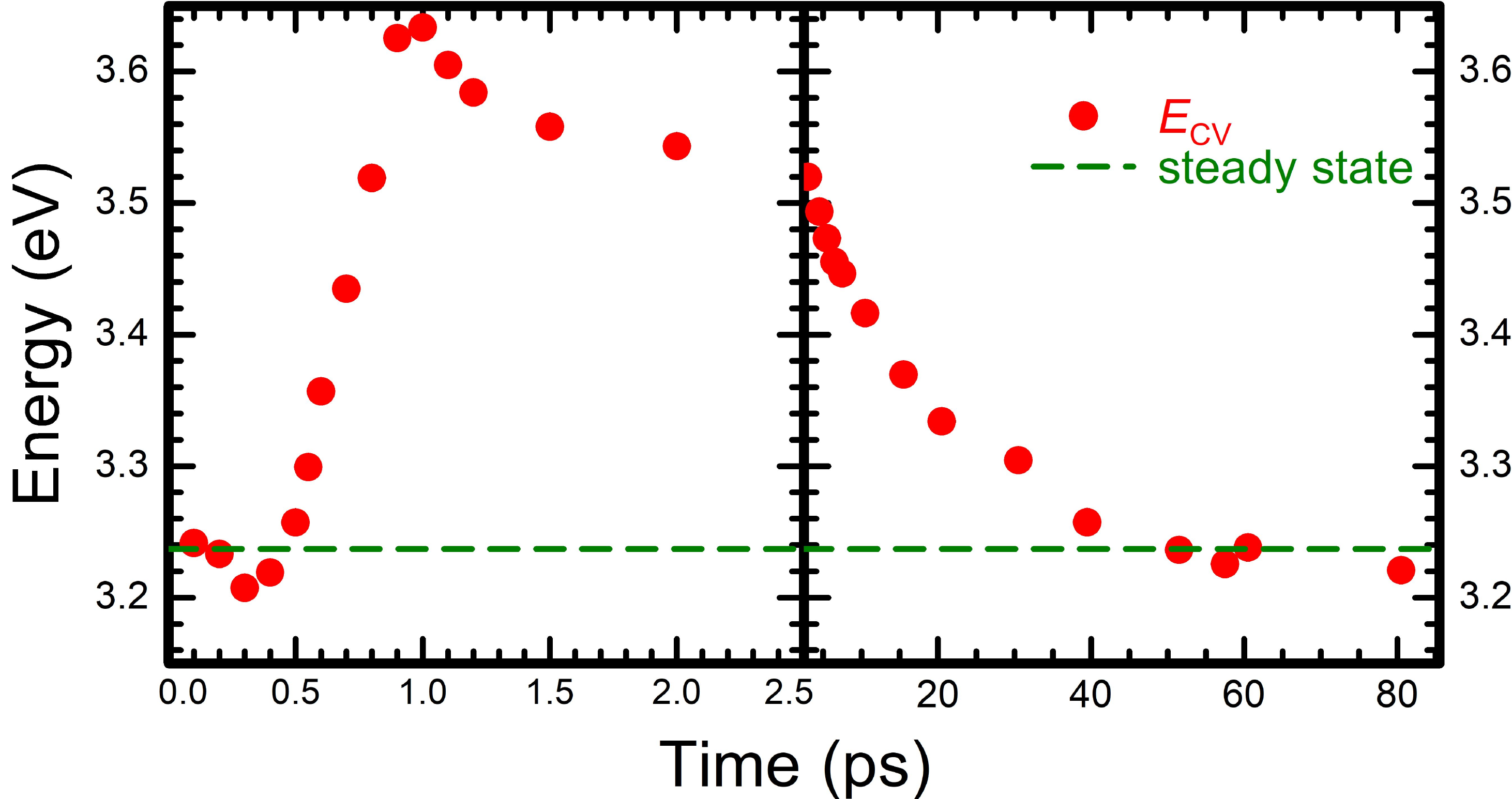}
\caption{Experimentally obtained transition energies $E_\mathrm{CV}$ for different times. Left: shortly after the pump beam, right: long after pump beam. The steady-state transition energy is indicated by the horizontal green dashed line.}
\label{fig_ECV_time}
\end{figure}

Next we take a look on the time-dependent transition energies for the excited GaN layer, which is at $x=0$ in the diffusion layer model. The obtained transition energies from the line shape-fit for each measurement are displayed in Fig. \ref{fig_ECV_time}. Here, an initial decrease of the transition energy is observed until $t=0.3$ps. Then, a sharp increase follows, until a turning point is reached around $t=1$ps. Obviously, for this point in time the relaxation is balanced by recombination and diffusion which indicates the maximum number of free charge carriers contributing to the shift of the absorption onset. For later times, the transition energy slowly returns to the steady-state case and even slightly drops below that value, again due to a stronger BGR contribution at this point, compared to the BMS contribution.

\begin{figure}[htb]
\includegraphics[angle=0, trim = 0mm 0mm 0mm 0mm, width=\columnwidth]{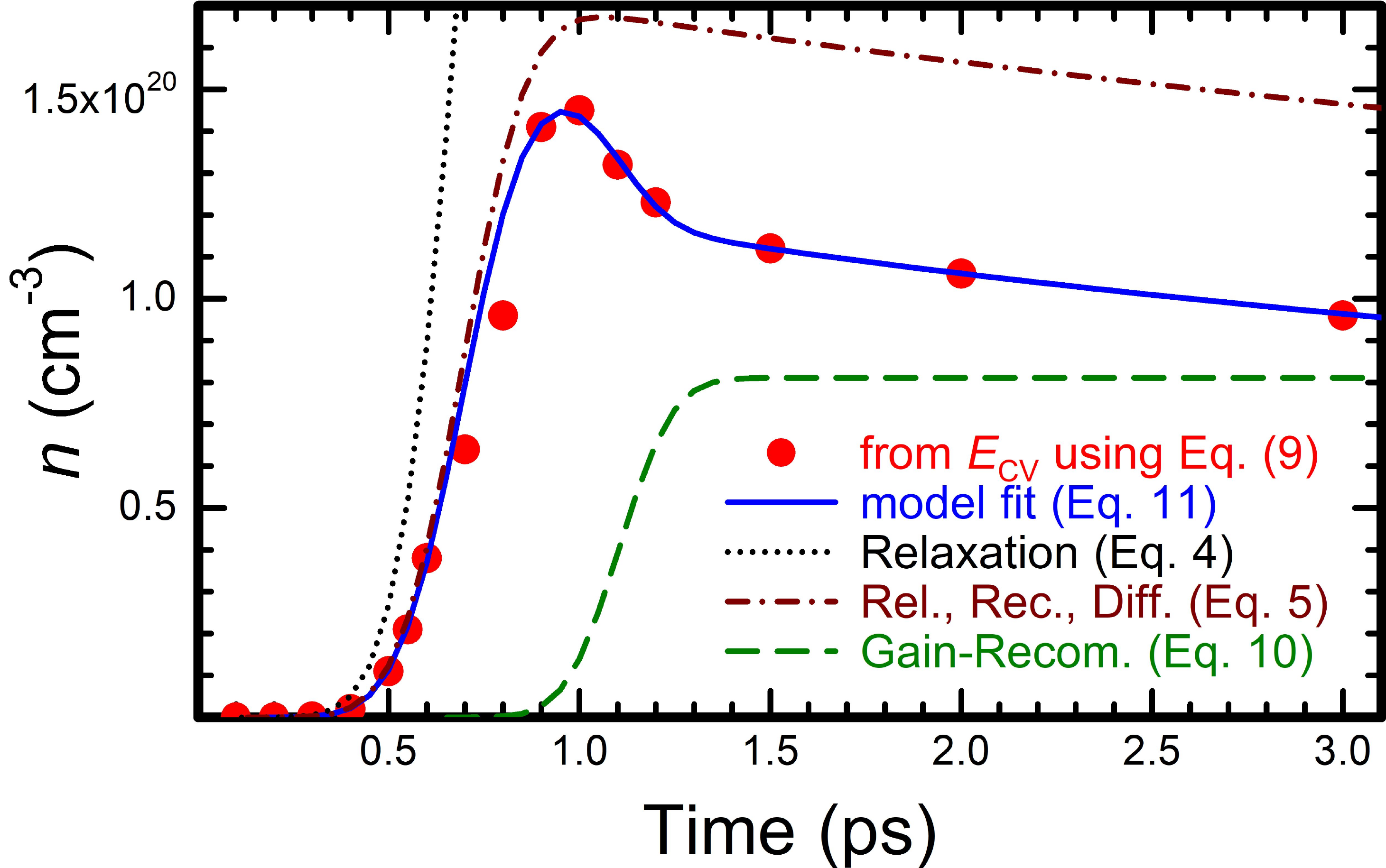}
\caption{Determined (red dots) and calculated (blue, continuous) free-carrier concentration in the CBM at the sample surface. The relaxation effect (black, dotted) as well as the combination of relaxation, recombination (without the gain effect) and diffusion (brown, dash-dotted) are displayed as well. Additionally, the free-carrier concentration 'lost' by the gain recombination (green, dashed) is included for better understanding.}
\label{fig_n_time_close}
\end{figure}

\begin{figure}[htb]
\includegraphics[angle=0, trim = 0mm 0mm 0mm 0mm, width=\columnwidth]{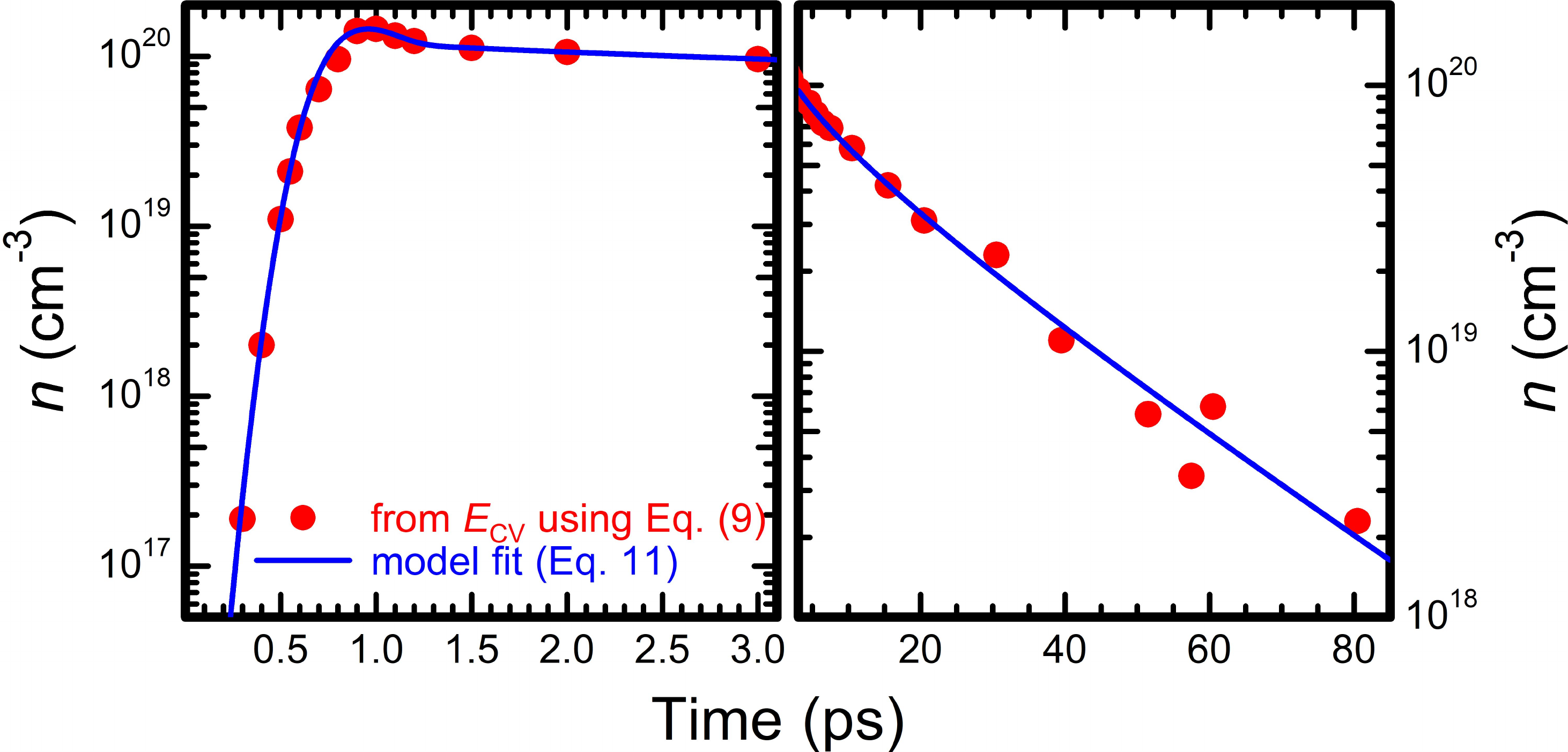}
\caption{Determined (red dots) and calculated (blue, continuous) free-carrier concentration in the CBM at the sample surface for different time scales. Left: shortly after the pump beam, right: long after pump beam (different $n$ axis).}
\label{fig_n_time}
\end{figure}

Using the approximations detailed in Sec. \ref{sec_band}, we convert the transition energies from Fig. \ref{fig_ECV_time} into free-electron concentrations at the CBM. The results are displayed in Figs. \ref{fig_n_time_close} and \ref{fig_n_time}. At $t\approx$1ps the free-carrier concentration in the CBM strongly decreases for only 0.3ps. We attribute this fast recombination effect to be different to the recombination dominating the remaining time region until the steady-state case is reached. Since this fast recombination process is limited to the spectra in which we included gain (see again Fig. \ref{fig_DF_short}), we label this effect as gain-recombination. The fact that this strong decrease only takes place in a short time period suggests, that this recombination process is not active the entire time but only for certain conditions. We assume the conditions to be dependent on the band-filling level. For instance, the gain-recombination only effects electrons over a defined energy or $k$-vector within the conduction band. We model the number of electrons recombining by gain as an error-function:

\begin{equation}
n_\mathrm{gain}\left(t\right)=\frac{N_2}{2}\cdot \left(\mathrm{erf}\left(\frac{t-\gamma_2}{\tau_2}\right)+1\right).
\label{eq_n_gain}
\end{equation} 

Here, $N_2$ and $\tau_2$ resemble the maximum number of recombined electrons by this effect and the characteristic recombination time, while $\gamma_2$ determines the position of the inflection point of the error function. To obtain the free-carrier trend, we subtract the gain-recombination from the relaxation process and then perform the recombination and diffusion model, as done in Eq. (\ref{eq_model_n}):

\begin{equation}
n_\mathrm{gm}\left(t\right)=\frac{\left(n_\mathrm{relax}-n_\mathrm{gain}\right)}{2}\cdot \mathrm{e}^{\alpha^2Dt-\frac{t}{\tau_1}}\cdot \mathrm{erfc}\left(\sqrt{\alpha^2Dt}\right).
\label{eq_model_gain}
\end{equation} 

The resulting model is also displayed in Fig. \ref{fig_n_time_close}. Looking at longer time scales (Fig. \ref{fig_n_time}), the free-carrier concentration steadily decreases exponentially due to recombination and diffusion. 

The model for the free-carrier concentration in the CBM at the sample surface in Eq. (\ref{eq_model_gain}) is fitted onto the determined free-carrier concentrations. The fit-results are presented in Tab. \ref{tab_parameters_gain}. A characteristic relaxation time of $\tau_0=0.19$ps was found which is comparable with previous, theoretical estimates \cite{PSSB_216_1999}. The obtained recombination time $\tau_1$ is also similar to previous investigations \cite{APL_70_1997,JAP_126_2019}. The recombination time and diffusion coefficient yield a diffusion length $L_\mathrm{D}=\sqrt{D\cdot\tau_1}=58.8$nm. This is lower compared to the diffusion length found in wurtzite GaN ($\approx 90\mathrm{nm}$) \cite{JAP_117_2015,JAP_120_2016}.

\begin{table}[htb]
  \caption{Parameters of the relaxation (Relax., $N_0,\gamma_0,\tau_0$), recombination (Recom., $\tau_1$), diffusion (Diff., $D$) and gain-recombination (Gain, $N_2,\gamma_2,\tau_2$) contributions for the free-carrier model in Eq. (\ref{eq_model_gain}) obtained by the best fit on the determined free-carrier concentration in Fig. \ref{fig_n_time}.}
  \begin{tabular}{ccccccc}
    \hline \hline
     & Relax. & Recom. & Diff. & Gain \\
    \hline
    $N_{0,2}$ (cm$^{-3}$) & $3.8\times 10^{20}$ & - & - & $8.1\times 10^{19}$ \\
    $\gamma_{0,2}$ (ps) & $0.70$ & - & - & $1.10$ \\
    $\tau_{0,1,2}$ (ps) & $0.19$ & $26.1$ & - & $0.16$ \\
    $D$ (cm$^2$/s) & - & - & $1.33$ & - \\
		\hline \hline
  \end{tabular}
\label{tab_parameters_gain}
\end{table}

\subsection{Temperature effects}
\label{sec_temperature}

Nevertheless, a different approach for considering the strong decrease of the transition energy or the free-carrier concentration can be made. Up until now, we supposed the sample to be at room-temperature. However, it is reasonable to assume that the pump-beam heats the affected area of the sample. The electron relaxation process is a thermalization via electron-phonon interaction. The generated phonons increase the lattice temperature of the GaN-layer by phonon-phonon and phonon-electron interaction with bound electrons. This increased sample temperature reduces the fundamental band gap of the material \cite{PRB_52_1995,EPOS_45_2011,PRB_89_2014}. The heating rate should be directly proportional to the electron relaxation process. Therefore, we approximate the increase in temperature by a time function with the same characteristic behavior as the relaxation process. The cooling back to room-temperature is assumed to be exponential. Consequently, the time-dependent temperature model results to:

\begin{equation}
T\left(t\right)=\frac{\left(T_1-T_0\right)}{2}\cdot \mathrm{e}^{-\frac{t}{\tau_4}}\cdot \left(\mathrm{erf}\left(\frac{t-\gamma_3}{\tau_3}\right)+1\right)+T_0
\label{eq_temperature}
\end{equation} 

with room-temperature $T_0=300$K and the maximum temperature $T_1$. The exponential cooling is described by $\tau_4$, while $\gamma_3$ and $\tau_3$ resemble the inflection point position and the characteristic heating time of the error-function. We set $\tau_3=\tau_0$. However, $\gamma_3$ can and should be different from $\gamma_0$, because a delay between phonon creation through electron relaxation and band gap reduction by phonon-phonon and phonon-electron interaction seems reasonable. \\
The fundamental band gap as a function of temperature $E_\mathrm{G}\left(T\left(t\right)\right)$ is estimated by an extrapolation of the semi-empirical model introduced by P\"assler \cite{PRB_66_2002_085201}. The necessary parameters for zb-GaN have been reported earlier \cite{PRB_85_2012}. This temperature dependent band gap is introduced in the BGR calculation in Eq. (\ref{eq_bgr}). For the calculations made here, we keep effective masses and the momentum matrix element $P$ constant, i.e. independent on temperature. The transition energy in Eq. (\ref{eq_ecv}) can now be described by applying the diffusion model from Eq. (\ref{eq_model_n}) and the temperature dependent band gap from Eq. (\ref{eq_temperature}) as:

\begin{equation}
E_\mathrm{CV}\left(t\right)=E_\mathrm{CV}\left(T\left(t\right),n_\mathrm{model}\left(t\right)\right).
\label{eq_ECV_temp}
\end{equation} 

\begin{figure}[htb]
\includegraphics[angle=0, trim = 0mm 0mm 0mm 0mm, width=\columnwidth]{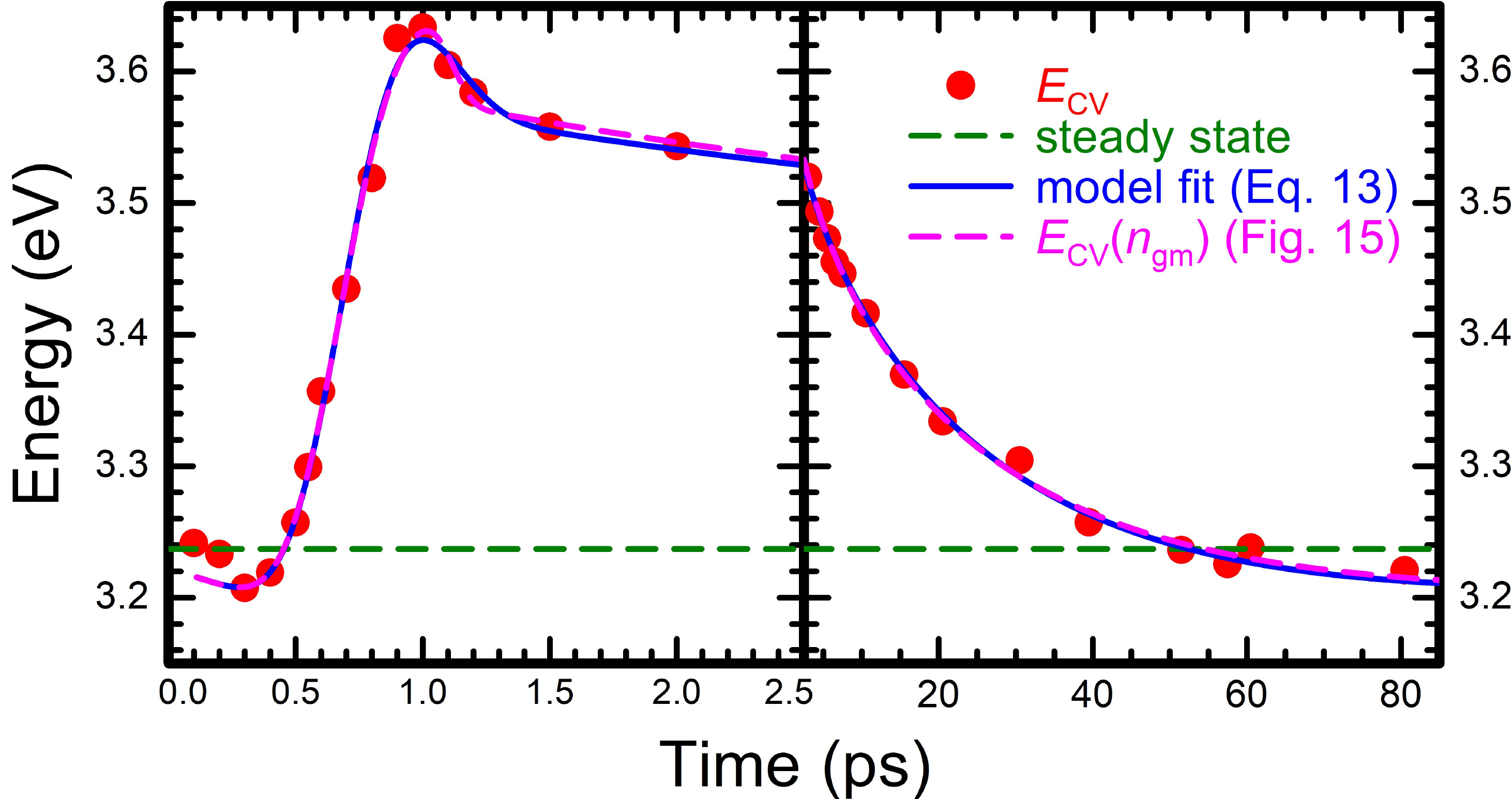}
\caption{Measured (red dots) and calculated by Eq. (\ref{eq_ECV_temp}) (blue, continuous) transition energies at the sample surface for different time scales. The results from Fig. \ref{fig_n_time} as transition energies are shown as well for comparison (pink, dashed). Left: shortly after the pump beam, right: long after pump beam.}
\label{fig_ECV_temp}
\end{figure}

This model transition energy is now fitted onto the experimental data from Fig. \ref{fig_ECV_time}. The fit-result is shown in Fig. \ref{fig_ECV_temp} together with the fit-result from Fig. \ref{fig_n_time} for comparison.

A good agreement between model and measurement data is achieved. The resulting parameters are gathered in Tab. \ref{tab_parameters_temp}. Fortunately, the values for relaxation, recombination and diffusion did not change dramatically compared to the alternative set of results (Tab. \ref{tab_parameters_gain}). However, this means that we can not differentiate between the possible gain recombination or a temperature increase. On one hand, the temperature increase through optical pumping must occur and therefore has to be considered. On the other hand, the measured imaginary part of the DF displays gain effects which can not be described solely by temperature increase. The fact, that we are able to model the behaviour of the transition energy exclusively using gain-recombination or temperature increase suggests that both effects contribute to the measured ultra-fast change of the absorption onset.

\begin{table}[htb]
  \caption{Best fit parameters of the relaxation (Relax., $N_0,\gamma_0,\tau_0$), recombination (Recom., $\tau_1$), diffusion (Diff., $D$), heating (Heat., $T_1,\gamma_3,\tau_3$) and cooling back to room-temperature (Cool., $\tau_4$) contributions for the transition energy model.}
  \begin{tabular}{cccccccc}
    \hline \hline
     & Relax. & Recom. & Diff. & Heat. & Cool. \\
    \hline
    $N_{0}$ (cm$^{-3}$) & $4.0\times 10^{20}$ & - & - & - & - \\
    $\gamma_{0,3}$ (ps) & $0.74$ & - & - & $1.09$ & - \\
    $\tau_{0,1,3,4}$ (ps) & $0.23$ & $23.6$ & - & $0.23$ & 19.13 \\
    $D$ (cm$^2$/s) & - & - & $0.76$ & - & - \\
    $T_{1}$ (K) & - & - & - & 543.54 & - \\
		\hline \hline
  \end{tabular}
\label{tab_parameters_temp}
\end{table}


\section{Summary}
\label{sec_summary}

In conclusion, the absorption onset of cubic GaN grown by MBE on a 3C-SiC/Si (001) substrate was investigated by trSE between -10ps and 5000ps before and after the pump-beam excitation respectively. The 266nm pump-beam induced a free-carrier profile in the GaN-layer which had to be considered in the data analysis by graded optical properties. The influence of absorption, recombination and diffusion processes yields different grading layers for different time scales. Then, the spectra around the band gap of the resulting pbp-DFs were described by a parametric model accounting for the band-band transition and potential gain effects, yielding the transition energies. These observed transition energies were then translated to free-carrier concentration by applying BMS and BGR models. The resulting time-dependent free-carrier concentration was described by a model containing relaxation, recombination, diffusion, and an additional gain-recombination effect, which yielded characteristic relaxation- and recombination-times as well as a diffusion coefficient. The change of the transition energy can alternatively be explained by considering heating of the sample. It is our presumption, that a mixture of both gain-recombination and heating (in combination with the previously mentioned effects) influence the absorption onset in zb-GaN after a high power pump-beam excitation.


\section*{Acknowledgment}
\label{sec_acknowledgment}

We gratefully acknowledge support by the Deutsche Forschungsgemeinschaft in the framework of Major Research Instrumentation Programs No. INST 272/230-1 and via project B02 within the Transregio program TRR 142 project number 231447078. 

We further acknowledge ELI Beamlines in Dolní Břežany, Czech Republic, for providing beamtime and thank the instrument group and facility staff for their assistance. This work was supported by the projects ADONIS (CZ.02.1.01/0.0/0.0/16-019/0000789) and ELIBIO (CZ.02.1.01/0.0/0.0/15-003/0000447) from the European Regional Development Fund, and by the project LM2018141 from the Czech Ministry of Education, Youth and Sport.


\end{document}